\documentclass[final,onefignum,onetabnum]{siamonline190516}

\usepackage{lipsum}
\usepackage{amsfonts}
\usepackage{graphicx}
\usepackage{epstopdf}
\usepackage{algorithmic}
\ifpdf
  \DeclareGraphicsExtensions{.eps,.pdf,.png,.jpg}
\else
  \DeclareGraphicsExtensions{.eps}
\fi

\usepackage{enumitem}
\setlist[enumerate]{leftmargin=.5in}
\setlist[itemize]{leftmargin=.5in}

\usepackage{caption}
\usepackage{subcaption}
\usepackage{bbm}

\newcommand{\E}{{\mathbb{E}}}
\newcommand{\R}{{\mathbb{R}}}
\newcommand{\bbP}{{\mathbb{P}}}
\newcommand{\N}{{\mathbb{N}}}
\newcommand{\bbI}{{\mathbbm{1}}}

\newsiamremark{remark}{Remark}
\newsiamremark{hypothesis}{Hypothesis}
\crefname{hypothesis}{Hypothesis}{Hypotheses}
\newsiamthm{claim}{Claim}

\headers{Entropic Hyper-Connectomes Computation and Analysis}{Michael G. Rawson}

\title{Entropic Hyper-Connectomes Computation and Analysis
}

\author{Michael G. Rawson\thanks{Department of Mathematics, University of Maryland at College Park, Maryland, USA 
  (\email{rawson@umd.edu}) and the Pacific Northwest National Laboratory}}

\usepackage{amsopn}

\begin{document}

\maketitle

\begin{abstract}
    Brain function and connectivity is a pressing mystery in medicine related to many diseases. Neural connectomes have been studied as graphs with graph theory methods including topological methods. Work has started on hypergraph models and methods where the geometry and topology is significantly different. We define a hypergraph called the hyper-connectome with joint information entropy and total correlation. We give the pseudocode for computation from finite samples. We give the theoretic importance of this generalization's topology and geometry with respect to random variables and then prove the hypergraph can be necessary for prediction and classification. We confirm with a simulation study and computation. We prove the approximation for continuous random variables with finite samples. We compare connectome versus hyper-connectome for predicting schizophrenia in subjects based on a fMRI dataset using a linear support vector machine. The hyper-connectome achieves better performance in accuracy (up to 56\%) and F1 score (up to 0.52) than the connectome. We reject null hypothesis at 95\% with p-value = 0.00074.
\end{abstract}

  



\section{Introduction}


Functional magnetic resonance imaging (fMRI) since the 1990's has been used to investigate brain activity in a benign and noninvasive way \cite{huettel2004}. fMRI detects oxygen usage by neurons with changing magnetic fields. fMRI is not sensitive enough to measure something as small as a neuron so a region of around 4mm by 4mm by 4mm is scanned which we call a \emph{voxel}. 
A connectome is a graph describing the connections between neurons. By grouping neurons into voxels (3D cubes in space), a graph can be constructed from fMRI scan data. Each vertex of the graph is a voxel and each weighted edge is the correlation between the voxels.
Correlation between voxels imply so-called functional connectivity, which is the statistical relationship between specific physiological signals in time\cite{EICKHOFF2015}.
This also assumes that neural activity is probabilistic. When the number of samples or measurements is low, the statistical confidence is low. To improve the confidence, voxels are often grouped into brain regions. Then each brain region has many samples and the confidence of the correlation is much greater. Additionally, the graph where the vertices correspond to just 61 brain regions is much smaller allowing for more computationally intensive graph analysis methods and this is what we shall use in \cref{sec:classification}. 

Brain fMRI scan data can be viewed as a hypergraph by using high order statistical methods. 
A hypergraph is a graph with edges between a tuple/group of vertices ($\ge 2$). We'll define a hypergraph model and calculate weights on the hyperedges. Hypergraph analyses can yield a deeper, more informative analysis on brain connectomes. We build a brain connectome hypergraph from fMRI scans to identify indicators that are impossible to retrieve from a standard brain connectome graph. Note that the hyper-connectome is \emph{not} a simplicial complex and thus most topological data analysis (TDA) methods, such as persistent homology, will not work here \cite{Stolz2021}. However, completing a hypergraph to a simplicial complex is another technique to consider. 

\section{Related Work}

There has been a large amount of activity around hypergraphs and their connection to connectomics in the recent past.
In the 2000s, Darling and Norris studied large random hypergraphs \cite{Darling2005} which is exactly what a hyper-connectome is. 
In 2018, hypergraph methods were compared and contrasted to TDA, sheaf methods, point cloud methods, and others by Purvine et al. \cite{Purvine2018}. 
Later, Aksoy et al. performed a general comparison between graphs and hypergraphs and their methods \cite{Aksoy2020}. 
On the connectomics front, in 2016, Munsell, Zu, Giusti, et al. considered different types of hypergraphs related to connectomes and neural data and various diseases \cite{Giusti2016, Munsell2016, Zu2016}. Sparse linear regression has been used to predict hyperedges in hypergraphs and classify disease \cite{Guo2018, Jie2016, Li2019}. 
Later, Sizemore et al. created a structural hypergraph of a mouse connectome \cite{Sizemore2019} and analyzed the topology. 
On the other hand, others learned, or optimized, a hypergraph as opposed to direct calculation \cite{Xiao2020, Zu2019}. 
Banka et al. learned autoencoder embeddings by hypergraphs \cite{Banka2020}. 
More recently, Stolz et al. analyzed the topology of connectomes in schizophrenic subjects versus normal subjects (siblings and non-siblings) \cite{Stolz2021}.

\section{Theory} 
Given two random variables, the Pearson correlation coefficient is \\
$ corr(X_1,X_2) := \frac{\E[(X_1 - \E X_1)(X_2 - \E X_2)]}
                        {\sqrt{\E((X_1-\E X_1)^2)}\sqrt{\E((X_2-\E X_2)^2)}} $.
This can be approximated given samples from distributions. The sample Pearson correlation coefficient, with $n$ samples, is 
$ corr(\hat X_1, \hat X_2) := \frac{\sum_i \frac{1}{n} (\hat X_{1,i} - \frac{1}{n} \sum_k \hat X_{1,k}) (\hat X_{2,i} - \frac{1}{n}\sum_k \hat X_{2,k})}
    {\sqrt{[\sum_i \frac{1}{n} (\hat X_{1,i} - \frac{1}{n}\sum_k \hat X_{1,k})] 
           [\sum_j \frac{1}{n} (\hat X_{2,j} - \frac{1}{n}\sum_k \hat X_{2,k})]}} $.
We will use this on fMRI samples to identify structure between brain regions. However, there are other statistics that we can use. The information entropy of a discrete random variable is 
$ H(X) = - \sum_{x\in\R} p(x) \log p(x) $ where $p$ is the probability that $X=x$. Note that we use the convention that $0 \cdot \log(0) = 0$. So again this can be approximated with samples. The sample information entropy, with $n$ samples, is 
$ H(\hat X) = - \sum_{x\in\R} \frac{1}{n}\sum_i\bbI_{\hat X_i = x}(\hat X_i) 
                                \log \left[\frac{1}{n}\sum_i\bbI_{\hat X_i = x}(\hat X_i) \right]
$. Now for any collection of discrete random variables, we can define the joint information entropy as
$ H(X_1,...,X_n) = - \sum_{x_1,...,x_n\in\R} p(x_1,...,x_n) \log p(x_1,...,x_n) $. The joint information entropy can be very useful for approximating how independent brain regions are versus how closely they collaborate in a predictable way. This is captured more directly by comparing the sum of entropies with the joint entropy. This difference is the mutual information or total correlation
\begin{align*}
    C(X_1,&X_2,...,X_n) := [\sum_i H(X_i)] - H(X_1,...,X_n).
\end{align*}
We use the following well known simplification. 
\begin{proposition}\label{prop:total_corr}
\begin{align}\label{eq:total_corr}
C(X_1,X_2,...,X_n) = \sum_{x_1,x_2,...,x_n \in \R} p(x_1,x_2,...,x_n) \log\frac{p(x_1,x_2,...,x_n)}{p(x_1)p(x_2)...p(x_n)}.
\end{align}
\end{proposition}

Proof in supplemental. Next we show how discrete random variables can approximate absolutely continuous random variables. 

\begin{theorem} \label{thm}
Let $Y_i$ be absolutely continuous random variables, $Y_i:\Omega\rightarrow\R$ measurable. 
The joint density $p_{Y}$ can be approximated by simple functions arbitrarily close in integration. We use an approximation and we say $p_{\tilde Y_i}$ is a simple function. 
Set discrete random variables $X_i$ to have a dirac for each term in simple function $p_{\tilde Y}$ with the coefficient so that the measures are the same:
$$ p_{\tilde Y} = \sum_k \alpha_k \bbI_{\Pi_i (a_{i,k},b_{i,k})},  
\quad p_{X} = \sum_k \beta_k \delta_{\Pi_i (b_{i,k}-a_{i,k})/2}, 
\quad \beta_k = \alpha_k \bbP_{\tilde Y}(\Pi_i (a_{i,k},b_{i,k})) $$

For any $\epsilon>0$, there is an discrete approximation with the total correlation
\begin{align*}
|C(&Y_1,Y_2,...,Y_n) - C(X_1,X_2,...,X_n)| < \epsilon. 
\end{align*}

\end{theorem}

Proof in supplemental. With the justification of \cref{thm}, we will approximate the total correlation from finite samples with \cref{alg:total_corr}. \cref{alg:total_corr} takes finite samples and approximates each probability in \eqref{eq:total_corr} with a localized mean of samples. 

\section{Simulation Study Results}\label{ssec:sim} 
First we'll consider a small example where we know the distributions. We'll setup the distributions where doing classification by hypergraph is in theory possible, which we prove. For random variables $X$ and $Y$, let 
$ X_i \sim Bernoulli(\{-1,1\},1/2),$ $ 1 \le i \le 3 $
and
$ Y = [Y_1, Y_2, Y_3] = [X_1 X_2, X_2 X_3, X_3 X_1 ] $.
Then
$ \sigma_{Y_i}^2 = \E Y_i^2 = 1 $.
For $i \ne j$, let $k,u,$ and $v$ be such that $Y_i=X_k X_u$ and $Y_j = X_k X_v$. Then 
\begin{align} \label{eq:corr}
 corr(Y_i, Y_j) = \frac{\E [Y_i Y_j]}{\sigma_{Y_i} \sigma_{Y_j}} = 0.
\end{align}
Proof in supplemental. This implies that the connectome Pearson correlation graph will not distinguish subject $X$ from subject $Y$. However, the \emph{total correlation} of $Y$ is not 0. 
\begin{proposition}\label{prop:total_corr_not_zero}
\begin{align*}
C(Y_1,Y_2,Y_3) &= [\sum_i H(Y_i)] - H(Y_1,Y_2,Y_3) < 0.
\end{align*}
\end{proposition}
Proof in supplemental. 
In this case, a discriminator can distinguish subject $Y$ from subject $X$ with total correlation but would fail to distinguish when using Pearson correlations. If this was a neural connectome, only the hyper-connectome could classify the subjects, because the hypergraph additionally contains the 2-simplices (or triangle edges).

We calculated the hypergraph weights with \cref{alg:total_corr}. Then we classify the data with a linear support vector machine. We give the results in \cref{table:sim}. The graph does not contain the information needed to classify the subject as a member of $X$ versus $Y$. The hypergraph does contain the information and the classifier successfully distinguishes $X$ from $Y$.

\begin{table}[h]
{\footnotesize
  \caption{Linear support vector machine classification of $X$ versus $Y$ from \cref{ssec:sim}. Hypergraph threshold $\epsilon = 10^{-5}$, dimension $d=3$, and variable count 3. Subjects are 1000 subject from $X$ and 1000 subjects from $Y$. There are 20 samples of each subject. The training/testing split is random 50\%.}  \label{table:sim}
\begin{center}
  \begin{tabular}{|c|c|c|c|} \hline
                & \bf Training Accuracy & \bf Testing Accuracy & \bf F1 Score \\ \hline
    Graph       & 51\% & 49\% & 0.66 \\ \hline
    Hypergraph  & 100\% & 100\% & 1 \\ \hline
  \end{tabular}
\end{center}
}
\end{table}

\section{Schizophrenia Dataset Results}

We next compute the hyper-connectome on real data. We utilize a schizophrenia (schiz.) fMRI dataset, see \cite{Wu2021}, consisting of 104 patients with schizophrenia and 124 healthy, normal controls. Between the groups, the age and gender differences are minimal (schiz.: age 36.88 $\pm$ 14.17 with 62 males, 41 females, 1 other, and 
Normal: age 33.75 $\pm$ 14.22 with 61 males and 63 females).
The fMRI aquisition details and preprocessing details are laid out in Adhikari et al. \cite{Adhikari2019}. This dataset consists of activity in 246 regions of interest (ROI) which we call the \emph{ROI variables} \cite{Fan2016}. Each ROI is formed from many distinct voxels. We create the connectome and hyper-connectome with a subset of these ROI variables. The vertices are each labelled with an ROI. 

\subsection{Hyper-Connectome Visualization} 
\begin{figure}[h]
     \centering
     \begin{subfigure}[b]{0.24\textwidth} 
         \centering
         \includegraphics[width=\textwidth]{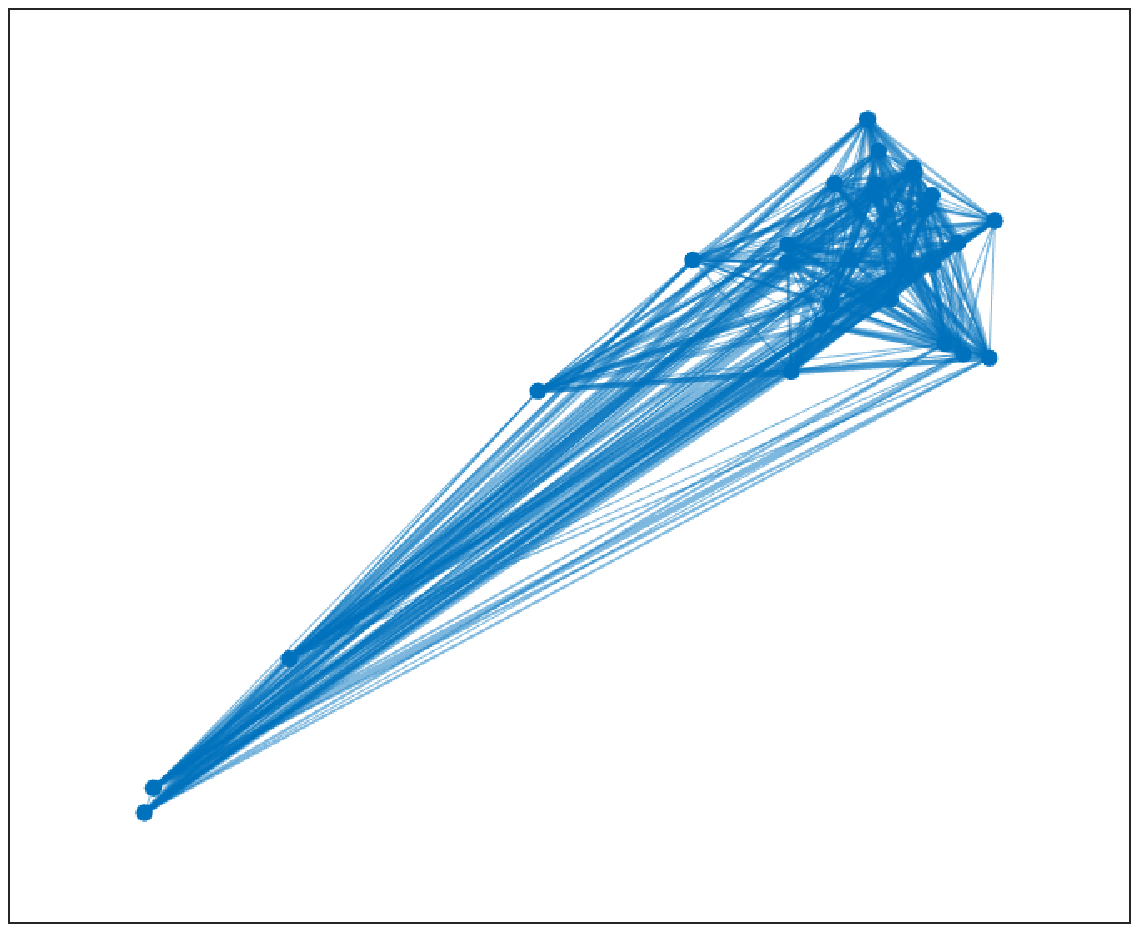}
         \caption{Normal Subject Connectome}
         \label{fig:1a}
     \end{subfigure}
     \hfill
     \begin{subfigure}[b]{0.24\textwidth}
         \centering
         \includegraphics[width=\textwidth]{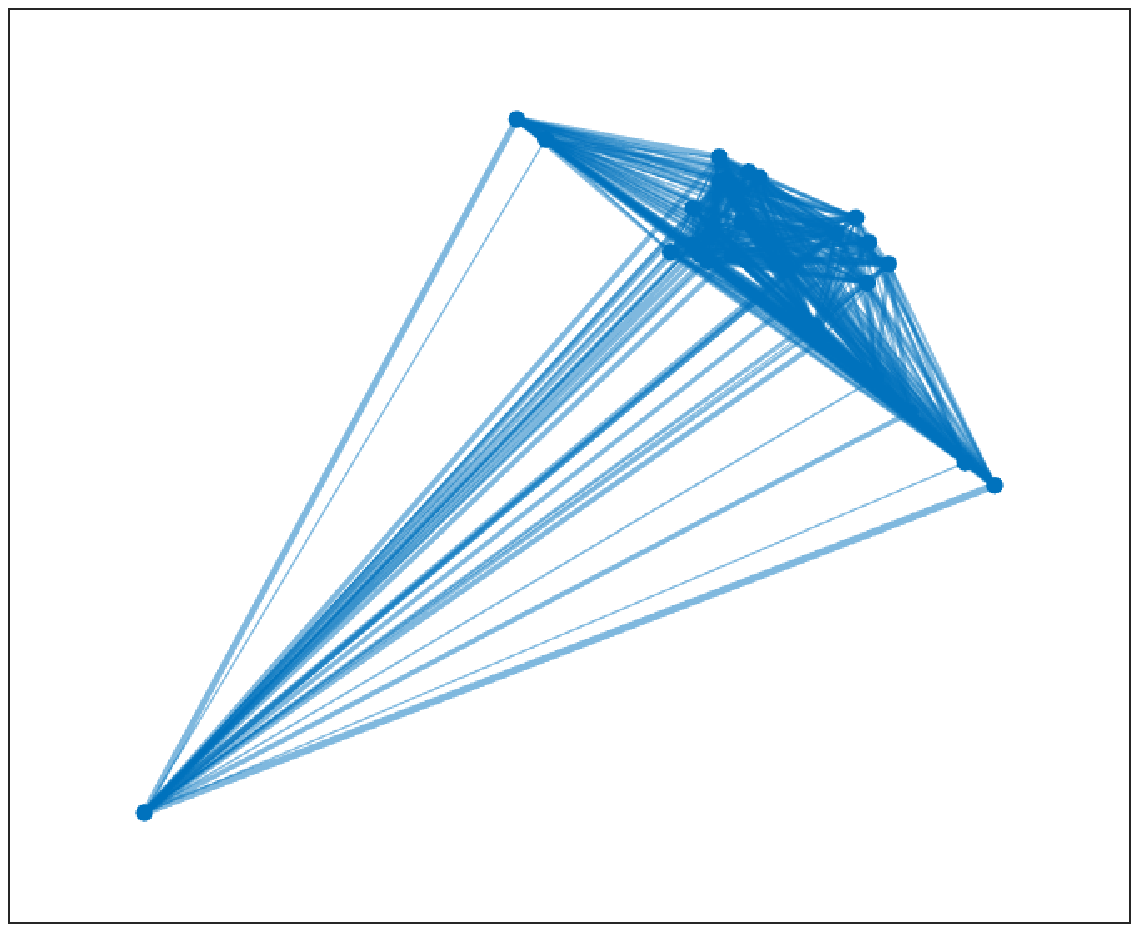}
         \caption{Schiz. Subject Connectome}
         \label{fig:1b}
     \end{subfigure}
     \hfill
     \begin{subfigure}[b]{0.24\textwidth}
         \centering
         \includegraphics[width=.8\textwidth]{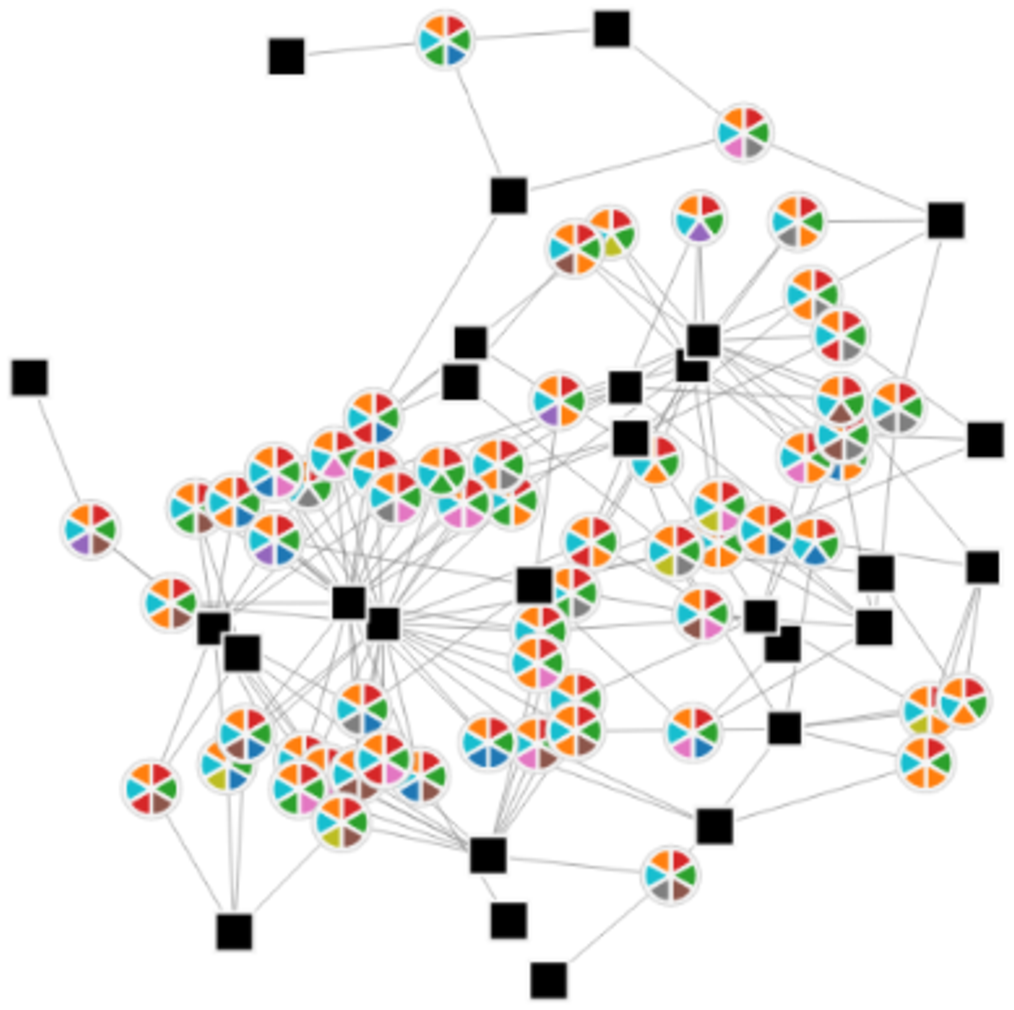}
         \caption{Normal Subject Hyper-Connectome}
         \label{fig:1c}
     \end{subfigure}
     \hfill
     \begin{subfigure}[b]{0.24\textwidth}
         \centering
         \includegraphics[width=.8\textwidth]{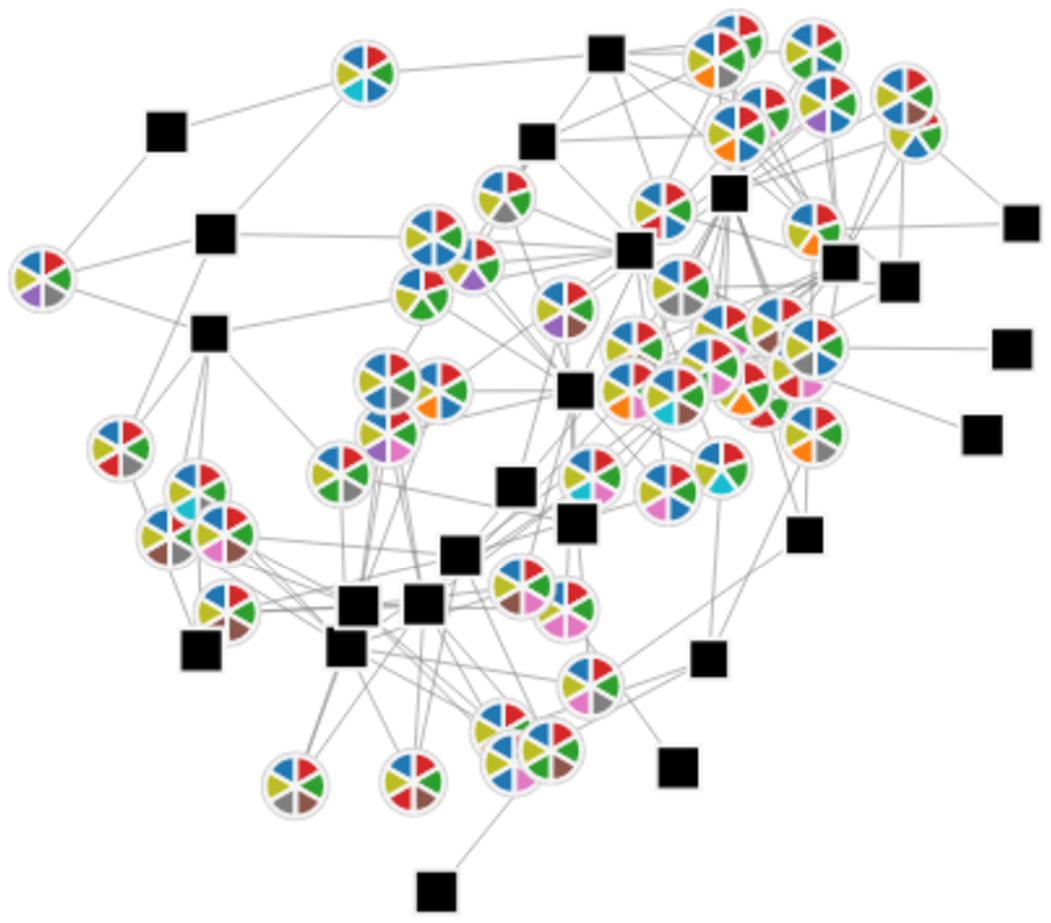}
         \caption{Schiz. Subject Hyper-Connectome}
         \label{fig:1d}
     \end{subfigure}
    \caption{(a) and (b): Connectome graph of a normal vs. a schiz. subject with 30 ROI nodes and correlation weighted edges. (c) and (d): Hyper-Connectome of a normal vs. a schiz. subject, plotting significant (weight $>2^8$) edges (circles) for 30 ROI nodes (squares). }
    \label{fig:1}
\end{figure}

We visualize connectomes by plotting the graph for a normal subject and schiz. subject in \cref{fig:1a} and \cref{fig:1b}. In this graph, the nodes are the brain regions and the edges are the absolute value of Pearson correlation between regions, where length and width indicate the weight of the edge. In \cref{fig:1c} and \cref{fig:1d}, we visualize the hyper-connectomes of the same two subjects. In this graph, the square nodes are the brain regions and the circles are significant hyperedges between multiple nodes \cite{zhou2021}. We see that the connectomes are highly clustered with few outliers versus the hyper-connectomes which are have nodes covering the connectedness spectrum. 

\begin{figure}[h]
     \centering
     \begin{subfigure}[t]{0.375\textwidth}
         \centering
         \includegraphics[width=\textwidth]{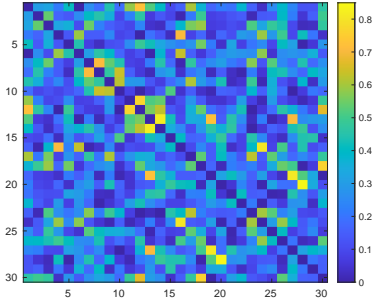}
         \caption{Normal Subject Pearson Correlation Matrix}
         \label{fig:2a}
     \end{subfigure}
     \hfill
     \begin{subfigure}[t]{0.375\textwidth}
         \centering
         \includegraphics[width=\textwidth]{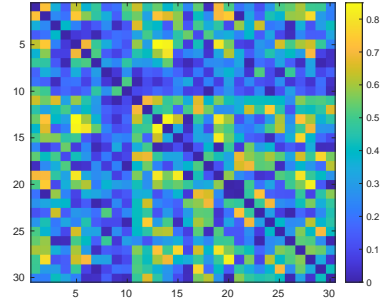}
         \caption{Schiz. Subject Pearson Correlation Matrix}
         \label{fig:2b}
     \end{subfigure}
     \hfill
     \begin{subfigure}[t]{0.375\textwidth}
         \centering
         \includegraphics[width=\textwidth]{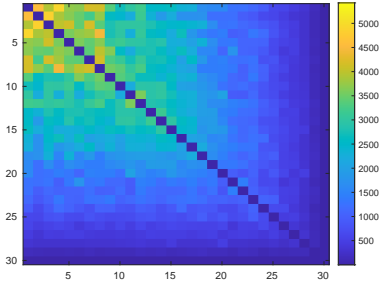}
         \caption{Normal Subject Total Correlation Tensor}
         \label{fig:2c}
     \end{subfigure}
     \hfill
     \begin{subfigure}[t]{0.375\textwidth}
         \centering
         \includegraphics[width=\textwidth]{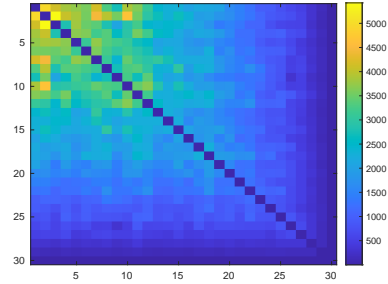}
         \caption{Schiz. Subject Total Correlation Tensor}
         \label{fig:2d}
     \end{subfigure}
    \caption{Top: Pearson correlation matrix of a normal vs. a schiz. subject, 30 ROIs. Bottom: Total correlation tensor of a normal vs. a schiz. subject, summed to matrix, 30 ROIs. }
    \label{fig:2}
\end{figure}

In \cref{fig:2}, we show the corresponding adjacency matrices to the graphs in \cref{fig:1} (ROI/vertices in same order). For each pair of ROI, we sum the weights of all common hyperedges to produce the adjacency matrices in \cref{fig:2c} and \cref{fig:2d}. We see that the magnitudes of \cref{fig:2a} and \cref{fig:2b} are similar (in 0 to 1) while the maximums of \cref{fig:2c} and \cref{fig:2d} differ significantly (5000 vs. 5500). We calculated the hypergraph with \cref{alg:total_corr}. The hypergraph threshold $\epsilon = 10^{-5}$, dimension $d=3$, and ROI variables are brain regions 1 to 30. The $\epsilon$ is chosen by searching and corresponds to sampling density. Samples used are the first 20 in the time series. 

\subsection{Hyper-Connectome Classification} \label{sec:classification} 
We showed in the previous section that the connectome and hyper-connectome can differ greatly in information content. The next question is how useful this is for distinguishing normal subjects and schizophrenic subjects. We vectorize the upper triangle of the connectome adjacency matrix and train a linear support vector machine to classify the subjects. After training, we calculate the accuracy and F1 score on the (unseen) test data. We follow the same procedure with the hyper-connectome. We report the results in \cref{table:real_data}. We find an increase in the testing accuracy of 6\% and in the F1 score of .08 from using the hyper-connectome versus the connectome. We compute the two-sample t-test and reject the equal testing accuracy mean hypothesis at 95\% confidence with p-value = 0.00074.

\begin{table}[h]
{\footnotesize
  \caption{Linear support vector machine prediction of schizophrenia. Hypergraph threshold $\epsilon = 10^{-5}$, dimension $d=3$, and ROI variables 1 to 61 considered. Samples over time are 20. The training/testing split is random 50\%. Calculations are average of 10 independent trials. Reject null hypothesis at 95\% with p-value = 0.00074. }  \label{table:real_data}
\begin{center}
  \begin{tabular}{|c|c|c|c|} \hline
                & \bf Training Accuracy & \bf Testing Accuracy & \bf F1 Score \\ \hline
    Graph       & 100\% & 50\% & 0.44 \\ \hline
    Hypergraph  & 100\% & 56\% & 0.52 \\ \hline
  \end{tabular}
\end{center}
}
\end{table}

\section{Conclusions} 
We have introduced the entropic hyper-connectome as a useful concept to study neuronal structure, function, and abnormalities. We have demonstrated this with fMRI data in vivo. From theory we have defined the hypergraph and proved that it can be necessary to detect various mixture distributions. We visualized the connectome and hyper-connectome and see significant differences. Finally, we trained a classifier to show that the hyperedges can improve classification with statistical significance.

\section*{Acknowledgments} 
The author thanks Dr. Shuo Chen, Dr. Elliot Hong, Dr. Peter Kochunov, the Maryland Psychiatric Research Center, and the University of Maryland School of Medicine for the fMRI schizophrenia dataset. The author thanks Dr. Michael Robinson for many helpful conversations. 

\bibliographystyle{siamplain}
\bibliography{references}

\newpage
\appendix
\section*{Supplementary Materials}

\begin{proof}[Proof of \cref{prop:total_corr}]
\begin{align*}
    C(&X_1,X_2,...,X_n)  = [\sum_i H(X_i)] - H(X_1,...,X_n) \\
                        &= [\sum_{x_1,...,x_n\in\R} p(x_1,...,x_n) \log p(x_1,...,x_n)] - [\sum_i \sum_{x_i} p(x_i) \log p(x_i)] \\
                        &= [\sum_{x_1,...,x_n\in\R} p(x_1,...,x_n) \log p(x_1,...,x_n)] - [\sum_i \sum_{x_i}\sum_{x_1,...,x_{i-1},x_{i+1},...,x_n\in\R} p(x_1,...,x_n) \log p(x_i)] \\
                        &= [\sum_{x_1,...,x_n\in\R} p(x_1,...,x_n) \log p(x_1,...,x_n)] - [\sum_{x_1,...,x_n\in\R} p(x_1,...,x_n) \sum_i \log p(x_i)] \\
                        &= \sum_{x_1,...,x_n\in\R} [p(x_1,...,x_n) \log p(x_1,...,x_n) - p(x_1,...,x_n) \sum_i \log p(x_i)] \\
                        &= \sum_{x_1,...,x_n\in\R} p(x_1,...,x_n) \log \frac{p(x_1,...,x_n)} {p(x_1)...p(x_n)}
\end{align*}
\end{proof}

\begin{proof}[Proof of \cref{prop:total_corr_not_zero}]
\begin{align*}
C(Y_1,Y_2,Y_3) &= [\sum_i H(Y_i)] - H(Y_1,Y_2,Y_3) 
\end{align*}
\begin{align*}
&= [\sum_i \bbP(Y_i=1) \log \bbP(Y_i=1) + \bbP(Y_i=-1) \log \bbP(Y_i=-1)] \\
&+ \bbP(Y_1=1,Y_2=1,Y_3=1) \log \bbP(Y_1=1,Y_2=1,Y_3=1) \\
&+ \bbP(Y_1=1,Y_2=1,Y_3=-1) \log \bbP(Y_1=1,Y_2=1,Y_3=-1)\\
&+ \bbP(Y_1=1,Y_2=-1,Y_3=1) \log \bbP(Y_1=1,Y_2=-1,Y_3=1)\\
&+ \bbP(Y_1=1,Y_2=-1,Y_3=-1) \log \bbP(Y_1=1,Y_2=-1,Y_3=-1)\\
&+ \bbP(Y_1=-1,Y_2=1,Y_3=1) \log \bbP(Y_1=-1,Y_2=1,Y_3=1) \\
&+ \bbP(Y_1=-1,Y_2=1,Y_3=-1) \log \bbP(Y_1=-1,Y_2=1,Y_3=-1)\\
&+ \bbP(Y_1=-1,Y_2=-1,Y_3=1) \log \bbP(Y_1=-1,Y_2=-1,Y_3=1)\\
&+ \bbP(Y_1=-1,Y_2=-1,Y_3=-1) \log \bbP(Y_1=-1,Y_2=-1,Y_3=-1)
\end{align*}
\begin{align*}
&= [\sum_i 1/2 \log 1/2 + 1/2 \log 1/2] \\
&+ \bbP(Y_1=1,Y_2=1,Y_3=1) \log \bbP(Y_1=1,Y_2=1,Y_3=1) \\
&+ \bbP(Y_1=1,Y_2=-1,Y_3=-1) \log \bbP(Y_1=1,Y_2=-1,Y_3=-1)\\
&+ \bbP(Y_1=-1,Y_2=1,Y_3=-1) \log \bbP(Y_1=-1,Y_2=1,Y_3=-1)\\
&+ \bbP(Y_1=-1,Y_2=-1,Y_3=1) \log \bbP(Y_1=-1,Y_2=-1,Y_3=1)\\
&= [3 \log 1/2] 
+ 1/4 \log 1/4 
+ 1/4 \log 1/4
+ 1/4 \log 1/4
+ 1/4 \log 1/4 \\
& = 3 \log 1/2 + \log 1/4 \\
& < 0
\end{align*}
\end{proof}

\begin{proof}[Proof of \cref{thm}]
Recall

$$ p_{\tilde Y} = \sum_k \alpha_k \bbI_{\Pi_i (a_{i,k},b_{i,k})},  
\quad p_{X} = \sum_k \beta_k \delta_{\Pi_i (b_{i,k}-a_{i,k})/2}, 
\quad \beta_k = \alpha_k \bbP_{\tilde Y}(\Pi_i (a_{i,k},b_{i,k})). $$

The total correlation
\begin{align*}
C(&\tilde Y_1,\tilde Y_2,...,\tilde Y_n) 
= \int_{y_1,y_2,...,y_n \in \R} p_{\tilde Y}(y_1,y_2,...,y_n) \log\frac{p_{\tilde Y}(y_1,y_2,...,y_n)}{p_{\tilde Y}(y_1)p_{\tilde Y}(y_2)...p_{\tilde Y}(y_n)}dy_1...dy_n.
\end{align*}
Write compactly as the integral of the simple function
$\Phi_{\tilde Y}(y)
= \sum_{k\in\N} \alpha_k \bbI_{\Pi_i (a_{i,k},b_{i,k})}(y)$

\begin{align*}
C(\tilde Y_1,\tilde Y_2,...,\tilde Y_n)
&= \int_{y \in \R^n} \Phi_{\tilde Y}(y) dy 
= \int_{y \in \R^n} \sum_{k\in\N} \alpha_k \bbI_{\Pi_i (a_{i,k},b_{i,k})}(y) dy \\
&= \sum_{k \in \N} \alpha_k \int_{\Pi_i (a_{i,k},b_{i,k})} dy \\
&= \sum_{k \in \N} \beta_k  \int \delta_{\Pi_i (b_{i,k}-a_{i,k})/2} \\
&= \sum_{x\in\R^n} \Phi_X(x) \\
 &= \sum_{x_1,x_2,...,x_n \in \R} p_X(x_1,x_2,...,x_n) \log\frac{p_X(x_1,x_2,...,x_n)}{p_X(x_1)p_X(x_2)...p_X(x_n)} \\
 &= C(X_1,X_2,...,X_n).
\end{align*}

\end{proof}

\begin{proof}[Proof of Equation \eqref{eq:corr}]
\begin{align*}
 &corr(Y_i, Y_j) = \frac{\E [Y_i Y_j]}{\sigma_{Y_i} \sigma_{Y_j}} \\
 &= \bbP(Y_i=1,Y_j=1) 
 - \bbP(Y_i=-1,Y_j=1) 
 - \bbP(Y_i=1,Y_j=-1) 
 + \bbP(Y_i=-1,Y_j=-1)\\
&=\bbP(X_1=1,X_2=1,X_3=1)+\bbP(X_1=-1,X_2=-1,X_3=-1)\\
 & \quad - 2\bbP(Y_i=-1,Y_j=1) 
+\bbP(X_k=-1,X_u=1,X_v=1)+\bbP(X_k=1,X_u=-1,X_v=-1)\\
&=1/8+1/8 - 2\bbP(X_k=-1,X_u=1,X_v=-1)-2\bbP(X_k=1,X_u=-1,X_v=1) + 1/8+1/8\\
&= 0
\end{align*}
\end{proof}

\begin{algorithm}[h]
\caption{Total Correlation}
\label{alg:total_corr}
\begin{algorithmic}
\STATE{Input:}
\STATE{\quad $M\in\N : $ variables}
\STATE{\quad $N\in\N : $ samples}
\STATE{\quad $d\in\N : $ dimensions}
\STATE{\quad $X\in\R^{M \times N} : $ measurements}
\STATE{\quad $\epsilon\in\R$ : threshold}
\STATE{Output:}
\STATE{\quad $T \in \R^{M^d}$ : total correlation}
\STATE{Begin:}
\STATE{$T=0 \in \R^{M^d}$}
\FOR{$1 \le i_1 \le M$}
\FOR{$i_1 \le i_2 \le M$}
\STATE{...}
\FOR{$i_{d-1} \le i_d \le M$}
\FOR{$1 \le s_1 \le N$}
\STATE{$\omega_1 = X(i_1,s_1)$}
\STATE{$p_1(\omega_1) = \frac{1}{N} \sum_j \bbI_{|X(i_1,j)-\omega_1|<\epsilon} (X)$}
\FOR{$1 \le s_2 \le N$}
\STATE{$\omega_2 = X(i_2,s_2)$}
\STATE{$p_2(\omega_2) = \frac{1}{N} \sum_j \bbI_{|X(i_2,j)-\omega_2|<\epsilon} (X)$}
\STATE{...}
\FOR{$1 \le s_d \le N$}
\STATE{$\omega_d = X(i_d,s_d)$}
\STATE{$p_d(\omega_d) = \frac{1}{N} \sum_j \bbI_{|X(i_d,j)=\omega_d|<\epsilon} (X)$}
\STATE{$p(\omega_1,\omega_2,...,\omega_d) = \frac{1}{N} \sum_j \bbI_{|X(i_1,j)-\omega_1|<\epsilon,...,|X(i_d,j)-\omega_d|<\epsilon}(X)$}
\STATE{$T(i_1,i_2,...,i_d) = T(i_1,i_2,...,i_d) + p(\omega_1,\omega_2,...,\omega_d) \log \frac{p(\omega_1,\omega_2,...,\omega_d)}{p_1(\omega_1)p_2(\omega_2)...p_d(\omega_d)}$}
\ENDFOR
\STATE{...}
\ENDFOR
\ENDFOR
\ENDFOR
\STATE{...}
\ENDFOR
\ENDFOR
\RETURN $T$
\end{algorithmic}
\end{algorithm}

\end{document}